\documentstyle[12pt]{article}
\newcommand{\bea}{\begin{eqnarray}}
\newcommand{\eea}{\end{eqnarray}}
\newcommand{\be}{\begin{equation}}
\newcommand{\ee}{\end{equation}}

\newcommand{\nn}{\nonumber}
\newcommand{\rf}[1]{(\ref{#1})}

\begin{document}
\begin{center}
{\Large\bf CONSTRAINED DYNAMICAL SYSTEMS: SEPARATION OF CONSTRAINTS INTO
FIRST AND SECOND CLASSES}\\
\vspace*{1cm}
N.P.Chitaia, S.A.Gogilidze\\
{\it Tbilisi State University, Tbilisi, University St.9, 380086 Georgia,} \\
\bigskip
and \\
\bigskip
Yu.S.Surovtsev \\
{\it Bogoliubov Laboratory of Theoretical Physics, Joint Institute for Nuclear
Research, Dubna 141 980, Moscow Region, Russia}
\end{center}
\begin{abstract}
In the Dirac approach to the generalized Hamiltonian formalism, dynamical
systems with first- and second-class constraints are investigated. The
classification and separation of constraints into the first- and second-class
ones are presented with the help of passing to an equivalent canonical set of
constraints. The general structure of second-class constraints is clarified.
\end{abstract}

PACS number(s): 02.20Tw, 03.65. Ge, 11.15. -q

\section{Introduction}
The generalized Hamiltonian formalism is a classical basis of the gauge
theories \cite{Dirac}. Originally the theories with the only first-class
constraints have played a main role among these theories, because the gauge
degrees of freedom are stipulated by the mentioned constraints. However, for
example, theories with the massive vector fields, supersymmetric and
superstring models introduce in consideration also the constraints of second
class. But the general case the generalized Hamiltonian dynamics, when in
addition to first-class constraints the second-class constraints are present
also in a theory, is studied relatively slightly up to now. Actually, there
are two approaches to treating constrained systems. In one approach, using the
classification of constraints into the first and second class, the
second-class constraints are disposed of by the Dirac brackets method
\cite{Dirac} - \cite{Henneaux-Teit}. Here the separation of constraints into
first- and second-class ones is needed, a possibility of which was indicated
by Dirac. But only in recent years there have appeared the real methods for
such separation \cite{Batlle-GPR} - \cite{Chaichian-Martinez-Lusanna},
developed, however, in the framework of the modified generalized Hamiltonian
formalism. In these papers, other than Dirac schemes were used for the
constraint proliferation, therefore, the question arises naturally about
equivalence of the constraint sets obtained in these works to the Dirac set.
Generally in the investigations of the dynamical systems with second-class
constraints there is a tendency (may be, not always justified) to modify
initial formulation of the generalized Hamiltonian dynamics
\cite{Batlle-GPR} - \cite{Henneaux-Slavnov}. Note that there is another more
recent approach \cite{Fadd-Jackiw} which does not apply the above
classification of constraints and where one has shown that for some Lagrangian
systems the basic bracket relations can be obtained without using the usual
Dirac brackets.

In this paper, we shall follow the conventional Dirac approach. Assuming a
complete set of constraints to be obtained according to the Dirac scheme for
breeding the constraints, we shall show that we can separate them into the
first- and second-class ones without modifying this scheme and solve the
problem of passing to an equivalent canonical set of constraints to
be used in a subsequent paper for deriving the local-symmetry transformation
generator and for proving the fact that second-class constraints do not
contribute to the law of these transformations unlike the assertions appeared
recently in the literature \cite{Sugano-Kimura} - \cite{Lusanna}.

\section{Classification and Separation of First- and Second-Class Constraints}

Here we restrict for simplicity ourselves to a system with a finite number of the
degrees of freedom $N$ described by a degenerate Lagrangian $L(q,\dot q)$,
where $q=(q_1,\cdots,q_N)$ and $\dot q=dq/dt=(\dot q_1,\cdots,\dot q_N)$
are generalized coordinates and velocities, respectively (all subsequent
considerations may be extended to the field theory by a standard way). After
passing to the Hamiltonian formalism, let $A$ primary constraints be obtained
in the phase space $(q,p)$. Further, provided that equations of motion are
self-consistent, all constraints are established according to the Dirac
procedure of breeding the constraints. The primary constraints will be denoted
by $\phi_\alpha^1$, where $\alpha=1,\cdots,A$ and
\begin{equation}
N-A=\mbox{rank} \left\|\frac{\partial^2L}{\partial\dot q_i\partial\dot
 q_j} \right\|.
\end{equation}
According to the Dirac scheme, from the stationary-state condition of the
primary constraints ~$\dot{\phi}_\alpha^1=\{\phi_\alpha^1,H_T\}=0$~($H_T$ is
the total Hamiltonian) we obtain the secondary constraints, denoted by
$\phi_\alpha^2$, from the stationary-state condition of which the tertiary
constraints $\phi_\alpha^3$ are obtained, etc. For consistent theories this
procedure is finished after the definite number of steps $M_\alpha-1$,  i.e.
~$\dot{\phi}_\alpha^{M_\alpha}=0$~ with taking account of all the previous
constraints. So, we have a system of constraints $\phi_\alpha^{m_\alpha}$,
where $\alpha=1,\cdots,A$ and $m_\alpha=1,2,\cdots,M_\alpha~~
(\sum_{\alpha=1}^{A}M_\alpha=M).$ The set of constraints
$\phi_\alpha^{m_\alpha}$ is complete and irreducible \cite{Dirac}.
Furthermore, let
\be \label{eq:rank-2R}
\mbox{rank} \left\|\{\phi_\alpha^{m_\alpha},\phi_\beta^{m_\beta}\}\right\|
=2R<M,
\ee
which implies the presence of $2R$ constraints of second class $\Psi_a^{m_a}$
and $M-2R$ constraints of first class $\Phi_\alpha^{m_\alpha}$ subjected to
the relation:
\bea
& & \{\Phi_\alpha^{m_\alpha}, \Phi_\beta^{m_\beta}\} \stackrel{\Sigma}{=}
\{\Phi_\alpha^{m_\alpha}, \Psi_a^{m_a} \} \stackrel{\Sigma}{=} 0,
\label{eq:PB-Phi-Psi}\\
& & \{\Psi_a^{m_a}, \Psi_b^{m_b}\} \stackrel{\Sigma}{=} F_{a~~b}^{m_a m_b}
\neq 0 \label{eq:PB-Psi-Psi}
\eea
($ \stackrel{\Sigma}{=}$ means this equality to hold on the surface of
all constraints $\Sigma$). The constraint sets $(\Phi,\Psi)$ and
$\phi_\alpha^{m_\alpha}$ are related with each other by the equivalence
transformation:
$$\left( \begin{array}{c}
{\bf \Phi} \\
{} \\
{\bf \Psi}
\end{array} \right) = {\bf S}
\left( \begin{array}{c}
{} \\
{\bf \phi} \\
{}
\end{array} \right),\qquad \mbox{det}{\bf S}\stackrel{\Sigma}{\neq} 0. $$
A possibility of constructing the set $(\Phi,\Psi)$ was indicated by Dirac.
However, for practical aims (for example, to elucidate a role of second-class
constraints in gauge transformations \cite{GSST}) the explicit form of set
$(\Phi,\Psi)$ is to be known. In what follows we shall busy ourself to obtain
this set of constraints through several successive stages being remain in the
framework of the Dirac approach and having in mind its employment next for
deriving the local-symmetry transformation generator.

Let us consider the antisymmetric matrix ${\bf K}^{11}$ with elements
$K_{\alpha\beta}^{11}=\{\phi_\alpha^1,\phi_\beta^1\}$, and let
\be \label{eq:rank-A_1}
\mbox{rank} \left\|K_{\alpha\beta}^{11}\right\|~\stackrel{\Sigma_1}{=}A_1=2R_1<A
\ee
($\Sigma_1$ is a primary constraint surface), i.e. $A_1$ primary constraints
exhibit their nature of second class already at this stage (more exactly, they
are candidates for this role provided that we shall be able to develop the
following procedure). One can regard the principal minor of rank $A_1$,
disposed in the left upper corner of the matrix ${\bf K}^{11}$, to be not
equal to zero. Write down
\be \label{eq:PB-phi^1-phi^1}
\{\phi_\alpha^1,\phi_\beta^1\}=f_{\alpha\beta\gamma}~\phi_\gamma^1+
D_{\alpha\beta},\quad \alpha,\beta,\gamma=1,\cdots,A
\ee
where $$D_{\alpha\beta}\stackrel{\Sigma_1}{=}F_{\alpha\beta}.$$
Among $F_{1\alpha}~(\alpha=2,\cdots,A_1)$ at least one element is non-zero in
accordance with the supposition \rf{eq:rank-A_1}. Renumbering the constraints one
can always obtain that $F_{12}\neq 0$.

Pass to a new set of constraints:
\bea \label{eq:phi-to-new1-phi}
\left. \begin{array}{l}
\;^1\phi_1^1\;=\;\phi_1^1\;,\qquad \;^1\phi_2^1\;=\;\phi_2^1\;, \\
%\quad
\;^1\phi_\alpha^1\;=\;\phi_\alpha^1\;+\;^1u_{\alpha 1}\;\phi_1^1~+\;^1u_
{\alpha 2}\;\phi_2^1~,\quad~~\alpha=3,\cdots,A.
\end{array}  \right.
\eea
The left superscripts indicate a stage of our procedure and will be omitted
in the resultant expressions. Determine the coefficients $\;^1u_{\alpha 1}\;$
and $\;^1u_{\alpha 2}\;$, which are functions of $q$ and $p$, by the following
expressions:
\be \label{eq:^1u}
\;^1u_{\alpha 1}\;=\;\frac{D_{2\alpha}}{D_{12}}\;,\quad
\;^1u_{\alpha 2}\;=\;-\frac{D_{1\alpha}}{D_{12}}\;,
\quad ~\alpha=3,4,\cdots,A,~~~
\ee
to guarantee the fulfilment of requirements:
\bea \label{eq:PB-^1phi-^1phi}
\;\{^1\phi_1^1\;,\;^1\phi_\alpha^1\}&=&f_{1\alpha\beta}~\phi_\beta^1\;+~D_
{1\alpha}~+~\{\phi_1^1~,\;^1u_{\alpha 1}\}\;\phi_1^1~+~f_{12\beta}{\;^1u}_
{\alpha 2}\;\phi_\beta^1~~ \nn\\
&&+~D_{12}\;{^1u}_{\alpha 2}\;+~\{\phi_1^1~,\;^1u_{\alpha 2}\}\;\phi_2^1\;
\stackrel{\Sigma_1}{=}\;0,\\
\;\{^1\phi_2^1\;,\;^1\phi_\alpha^1\}&=&f_{2\alpha\beta}~\phi_\beta^1\;+~D_
{2\alpha}~+~\{\phi_2^1~,\;^1u_{\alpha 1}\}\;\phi_1^1~+~f_{21\beta}{\;^1u}_
{\alpha 1}\;\phi_\beta^1~~ \nn\\
&&+~D_{21}{\;^1u}_{\alpha 1}\;+~\{\phi_2^1~,\;^1u_{\alpha 2}\}\;\phi_2^1\;
\stackrel{\Sigma_1}{=}\;0.\nn
\eea
With the help of \rf{eq:PB-phi^1-phi^1} and \rf{eq:^1u}, it is easily seen
that
$$^1D_{12}=\;-^1D_{21}\;\stackrel{\Sigma_1}{=}\;^1F_{12}\;=F_{12}~\neq 0\;,
\quad
^1D_{\alpha\beta}~\stackrel{\Sigma_1}{=}F_{\alpha\beta}~=0,\quad \alpha=1,2,~~
\beta=3,4,\cdots,A. $$
So, by means of the transformation
\be \label{eq:^1phi-Lambda-phi}
^1\phi_\alpha^1=\;^1\Lambda_{\alpha\beta}\;\phi_\beta^1,\qquad
\mbox{det}\|^1\Lambda_{\alpha\beta}\|=1,
\ee
\be \label{eq:^1Lambda}
\;^1{\bf \Lambda}\;=\left( \begin{array}{cc}
^1{\bf I}_1 & \;^1{\bf O}\; \\
^1{\bf U} & ^1{\bf I}_2
\end{array}  \right),
\ee
where $^1{\bf I}_1$, $^1{\bf I}_2$ and $^1{\bf O}$ are the unit $2\times 2$-,
$(A-2)\times (A-2)$- and zero $2\times (A-2)$-blocks, respectively, and
$$^1{\bf U}=\left( \begin{array}{cc}
^1u_{31} & ^1u_{32}  \\
\vdots & \vdots  \\
^1u_{A 1} & ^1u_{A 2}
\end{array}  \right), $$
we obtain at the first stage of our procedure:
\be \label{eq:^1K11}
^1K_{\alpha\beta}^{11}~=\;\{^1\phi_\alpha^1\;,\;^1\phi_\beta^1\}\;=
\;^1\Lambda_{\alpha\sigma}\;^1\Lambda_{\beta\tau}\;K_{\sigma\tau}^{11}\;+~
O(\phi_\alpha^1),
\ee
$$\;^1{\bf K}^{11}\;\stackrel{\Sigma_1}{=}\left( \begin{array}{cc}
F_{12}\cdot {\bf J} & {\bf O} \\
{\bf O} & \|^1{\cal F}_{\alpha\beta}\|(\alpha,\beta=3,4,\cdots,A)
\end{array}  \right), $$
where
${\bf J}=\left( \begin{array}{cc}
0 & 1 \\
-1 & 0
\end{array}  \right)$ and ${\bf O}$ are zero blocks, and  $\|^1{\cal F}_
{\alpha\beta}\|$ is $(A-2)\times (A-2)$-block, which must be reduced to the
quasidiagonal form at the next stages of procedure.

It is evident, we have
$$\mbox{rank} \left\|^1{\cal F}_{\alpha\beta}\right\|_{\alpha,\beta=
3,4,\cdots,A}=A_1-2,$$
therefore among elements $^1F_{3\beta}~ (\beta=4,\cdots,A_1)$ at least one
(let $^1F_{34}$) is not equal to zero. Repeating the above procedure in
respect to this block, i.e. making the transformation
\be \label{eq:^1phi-to-^2phi}
^2\phi_\alpha^1=
\left\{ \begin{array}{l}
^1\phi_\alpha^1~,\quad \alpha=1,2,3,4, \\
^1\phi_\alpha^1~+\;^2u_{\alpha 3}\;{^1\phi_3^1}\; +\;^2u_{\alpha 4}\;
{^1\phi_4^1}\;,\quad
\alpha=5,6,\cdots,A,
\end{array}  \right.
\ee
where functions $^2u_{\alpha 3}$ and $^2u_{\alpha 4}$ are determined as
\be \label{eq:^2u}
^2u_{\alpha 3}~=\frac{\;^1D_{4\alpha}\;}{\;^1D_{34}\;},\quad
^2u_{\alpha 4}~=-\frac{\;^1D_{3\alpha}\;}{\;^1D_{34}\;}, \quad
\alpha=5,6,\cdots,A,
\ee
we satisfy the requirements
\be \label{eq:PB-^2phi-^2phi}
\;\{^2\phi_3^1\;,\;^2\phi_\alpha^1\}\;\stackrel{\Sigma_1}{=}0,\qquad
\;\{^2\phi_4^1\;,\;^2\phi_\alpha^1\}\;\stackrel{\Sigma_1}{=}0.
\ee
One can see with the help of \rf{eq:PB-phi^1-phi^1}, \rf{eq:^2u} and
\rf{eq:PB-^2phi-^2phi} that
$$^2D_{34}~=\;-^2D_{43}\;\stackrel{\Sigma_1}{=}\;^1F_{34}\;\neq 0,
\quad
^2D_{\alpha\beta}~\stackrel{\Sigma_1}{=}\;^2F_{\alpha\beta}\;=0,\quad \alpha
=3,4,~~\beta=5,6,\cdots,A $$
and, furthermore,
$$\;^2D_{1\beta}\;=\;^1D_{1\beta}\;~(\beta=2,3,4),\quad \;^2D_{23}\;=
\;^1D_{23}\;,\quad ^2D_{24}~=\;^1D_{24}\;,$$
$$^2D_{\alpha\beta}~\stackrel{\Sigma_1}{=}\;^1D_{\alpha\beta}~~(\alpha=1,2,~~
\beta=5,6,\cdots,A),$$
i.e. the structure of zero blocks and the principal left minor, which is
obtained at the first stage, has survived.

So, at the second stage, with the help of the transformation
\be \label{eq:^2phi-Lambda-^1phi}
^2\phi_\alpha^1~=\;^2\Lambda_{\alpha\beta}\;{^1\phi_\beta^1}~=
\;^2\Lambda_{\alpha\beta}\;{^1\Lambda_{\beta\sigma}}~\phi_\sigma^1~,\qquad
\mbox{det}\|^2\Lambda_{\alpha\beta}~{^1\Lambda_{\beta\sigma}}\|=1,
\ee
\be \label{eq:^2Lambda}
^2{\bf \Lambda}~=\left( \begin{array}{cc}
^2{\bf I}_1 & ^2{\bf O} \\
^2{\bf U} & ^2{\bf I}_2
\end{array}  \right),
\ee
where $^2{\bf I}_1$, $^2{\bf I}_2$ and $^2{\bf O}$ are the unit $4\times 4$-,
$(A-4)\times (A-4)$- and zero $4\times (A-4)$-matrices, respectively, and
$$^2{\bf U}=\left( \begin{array}{cccc} 0 & 0 & ^2u_{53} & ^2u_{54} \\ \vdots &
\vdots & \vdots & \vdots \\ 0 & 0 & ^2u_{A 3} & ^2u_{A 4} \end{array}
\right),$$
we obtain
\bea \label{eq:^2K11} ^2{\bf K}^{11}~\stackrel{\Sigma_1}{=}
\left( \begin{array}{c|c} {\begin{array}{cc}
F_{12}\cdot {\bf J} & {\bf O} \\
{\bf O} & ^1F_{34}\cdot {\bf J}
\end{array}} & {\bf O}\\
\hline
{\bf O} & \|^2{\cal F}_{\alpha\beta}\|(\alpha,\beta=5,6,\cdots,A)
\end{array}  \right).
\eea
Iterating  now  this  procedure  ~$R_1=A_1/2$~  times, we  shall  receive  the
matrix ~$^{R_1}{\bf K}^{11}~=\\ \left\|\{\;^{R_1}\phi_\alpha^1\;,\;^{R_1}
\phi_\beta^1\;\} \right\|$~ in the quasidiagonal form on the
primary-constraint surface $\Sigma_1$:
\bea \label{eq:^R_1K11}
\;^{R_1}{\bf K}^{11}\;\stackrel{\Sigma_1}{=} \left(
\begin{array}{c|c} {\begin{array}{cccc} F_{12}\cdot {\bf J} & {\bf O} & \ldots
& {\bf O}\\ {\bf O} & ^1F_{34}\cdot {\bf J} & \ldots & {\bf O}\\ \vdots &
\vdots & \ddots & \vdots \\ {\bf O} & {\bf O} & \ldots &
\;^{R_1-1}F_{A_1-1~A_1}\cdot {\bf J}\; \end{array}} & {\bf O}\\ \hline {\bf O}
& {\bf O} \end{array}  \right).  \eea The corresponding equivalent set of
primary constraints is determined by the relation:  \be
\label{eq:^R_1phi-Lambda-^1phi} ^{R_1}\phi_\alpha^1~=\;^{R_1}\Lambda_
{\alpha\beta}\;{^{R_1-1}\Lambda_{\beta\gamma}}\cdots {^1\Lambda_{\sigma\tau}}~
{^1\phi_\tau^1}~=\overline{\Lambda}_{\alpha\beta}\;\phi_\beta^1~,\qquad
\mbox{det}\overline{{\bf \Lambda}}=1.
\ee
Among $A$ primary constraints, $A_1=2R_1$ functions had exhibited their nature
of second class already in interaction with each other. In described procedure
it is important that every subsequent stage preserves the structure of zero
blocks and principal left minor obtained at the preceding stage. We shall
denote the second-class constraints by the letter $\psi(\Psi)$. Thus the
following set of primary constraints is obtained:
\be \label{eq:psi1-phi1}
[\psi_{a_1}^1]_{a_1=1}^{A_1}~,\qquad [\phi_{\alpha_1}^1]_{\alpha_1=1}^{A-A_1}
\ee
with properties
\be \label{eq:PB-psi1-psi1}
\{\psi_{a_1}^1,\psi_{b_1}^1\}\stackrel{\Sigma_1}{=}\left\{\begin{array}{ll}
F_{a_1b_1}\neq 0,\quad & a_1=2k+1,~b_1=2k+2~~\mbox{and}\\
{}& \mbox{conversely}~(k=0,1,\cdots,A_1-2),\\
0,& \mbox{in other cases},
\end{array} \right.
\ee
\be \label{eq:PB-psi1-phi1}
\{\psi_{a_1}^1,\phi_{\alpha_1}^1\}~\stackrel{\Sigma_1}{=}0,\qquad
\{\phi_{\alpha_1}^1,\phi_{\beta_1}^1\}~\stackrel{\Sigma_1}{=}0.
\ee
It is clear that constraints $\psi_{a_1}^1$ do not generate secondary
constraints. Furthermore, one can attain that
\be \label{eq:PB-psi1-phi-second1}
\{\psi_{a_1}^1,\phi_{\alpha}^{m_\alpha}\}~\stackrel{\Sigma_1}{=}0,\qquad
m_\alpha=2,\cdots,M_\alpha.
\ee
To this end, we shall make the transformation
\be \label{eq:phi-second1-to-1phi-second1}
^1\phi_{\alpha}^{m_\alpha}~=\phi_{\alpha}^{m_\alpha}~+~C_{\alpha b_1}^
{m_\alpha}~\psi_{b_1}^1~.
\ee
Then using the definition
$$\{\psi_{a_1}^1,\phi_{\alpha}^{m_\alpha}\}~\stackrel{\Sigma_1}{=}
F_{a_1~\alpha}^{1~m_\alpha}$$
and taking account of \rf{eq:PB-psi1-psi1}, we shall meet the request
\rf{eq:PB-psi1-phi-second1} provided that
$$ C_{\alpha b_1}^{m_\alpha}~=-\frac{F_{a_1~\alpha}^{1~m_\alpha}}
{F_{a_1b_1}}~,$$
where if ~$a_1=2k+1$, then ~$b_1=2k+2$~ and conversely ($k=0,1,\cdots,A_1-2$).

Now let us turn to $\phi_{\alpha_1}^{m_{\alpha_1}},~\alpha_1=1,\cdots,A-A_1.$
Let
\be \label{eq:rank-A_2}
\mbox{rank}\left\|\{\phi_{\alpha_1}^1,\phi_{\beta_1}^2\}\right\|~
\stackrel{\Sigma}{=}A_2<A-A_1.
\ee
Furthermore, we have
\be \label{eq:phi1-phi2-sym}
\{\phi_{\alpha_1}^1,\phi_{\beta_1}^2\}~\stackrel{\Sigma}{=}
~\{\phi_{\beta_1}^1,\phi_{\alpha_1}^2\}.
\ee
In considering the matrix $\left\|\{\phi_{\alpha_1}^1,\phi_{\beta_1}^2\}
\right\|$ one can regard the principal minor of rank $A_2$, disposed in
the left upper corner of this matrix, to be not equal to zero. We denote it by
$${\bf K}^{12}=\left\|\{\phi_{a_2}^1,\phi_{b_2}^2\}\right\|~,~~~~ \mbox{where}
~~~a_2,b_2=1,\cdots,A_2.$$
Using the procedure which is analogous to the one for quasidiagonalization of
the matrix ${\bf K}^{11}$, we shall obtain the matrix ${\bf K}^{12}|_{\Sigma}$
in the diagonal form. To this end, we notice at first that
$\{\phi_1^1,\phi_1^2\}\stackrel{\Sigma}{\neq}0$. We make the transformation
\be \label{eq:phi^1-1phi^1}
\;^1\phi_1^1\;=\;\phi_1^1\;,\quad
\;^1\phi_a^1\;=\;\phi_a^1\;+\;^1u_{a 1}\;\phi_1^1~,\quad~~a=2,\cdots,A_2,
\ee
from here
\be \label{eq:phi^2-1phi^2}
\;^1\phi_1^2\;=\;\phi_1^2\;,\quad
\;^1\phi_a^2\;=\;\phi_a^2\;+\;^1u_{a 1}\;\phi_1^2~,\quad~~a=2,\cdots,A_2,
\ee
where $\;^1u_{a 1}\;$ is taken as
\be \label{eq:1u_a1}
\;^1u_{a 1}\;=-\frac{D_{1a}^{12}}{D_{11}^{12}}
\ee
to satisfy the requirement
$$\;\{^1\phi_1^1\;,\;^1\phi_a^2\}\;\stackrel{\Sigma}{=}
\;\{^1\phi_a^1\;,\;^1\phi_1^2\}\;\stackrel{\Sigma}{=}0.$$
Moreover, we have
\be \label{eq:PB-phi_2-phi_2}
\;\{^1\phi_2^1\;,\;^1\phi_2^2\}\;\stackrel{\Sigma}{=}\;^1F_{22}^{12}\;=
F_{22}^{12}-\frac{(F_{12}^{12})^2}{F_{11}^{12}}\neq 0.
\ee
Further making the transformation
\be \label{eq:1phi^1-2phi^1}
\;^2\phi_2^1\;=\;^1\phi_2^1\;,\quad
\;^2\phi_a^1\;=\;^1\phi_a^1\;+\;^2u_{a 2}~^1\phi_2^1\;,\quad~~a=3,\cdots,A_2,
\ee
and, therefore,
\be \label{eq:1phi^2-2phi^2}
\;^2\phi_2^2\;=\;^1\phi_2^2\;,\quad
\;^2\phi_a^2\;=\;^1\phi_a^2\;+\;^2u_{a 2}~^1\phi_2^2\;,\quad~~a=3,\cdots,A_2,
\ee
we determine $\;^2u_{a 2}\;$ as
\be \label{eq:2u_a2}
\;^2u_{a 2}\;=-\frac{\;^1D_{2a}^{12}\;}{\;^1D_{12}^{22}\;}
\ee
to satisfy the requirement
$$\;\{^2\phi_2^1\;,\;^2\phi_a^2\}\;\stackrel{\Sigma}{=}
\;\{^2\phi_a^2\;,\;^2\phi_2^2\}\;\stackrel{\Sigma}{=}0.$$
Furthermore, we have
\be \label{eq:PB-phi_3-phi_3}
\;\{^2\phi_3^1\;,\;^2\phi_3^2\}\;\stackrel{\Sigma}{=}\;^2F_{33}^{12}\;=
\;^1F_{33}^{12}\;-\frac{\;(^1F_{23}^{12})^2\;}{\;^1F_{22}^{12}\;}\neq 0.
\ee
So, we have obtained
$$^2{\bf K}^{12}~\stackrel{\Sigma}{=}\left( \begin{array}{c|c}
{\begin{array}{cc}
F_{11}^{12} & 0 \\
0 & \;^1F_{22}^{12} \end{array}} & {\bf O}\\
\hline
{\bf O} & \|^2{\cal F}_{ab}\|(a,b=3,4,\cdots,A_2)
\end{array}  \right).$$
Continuing this process, we shall deduce at the $(A_2-1)$ stage that with the
help of the equivalence transformation of those primary constraints which have
the nonvanishing Poisson brackets on $\Sigma$ with their secondary constraints
(the latter are obtained by the Dirac procedure), the matrix ${\bf K}^{12}|_
{\Sigma}$ is leaded to the diagonal form with the nonvanishing diagonal
elements
$$\{\psi_{a_2}^1,\psi_{a_2}^2\}~\stackrel{\Sigma}{=}~F_{a_2~a_2}^{1~2}\neq 0
~~~(a_2=1,\cdots,A_2).$$
Note, the constraints, which have exhibited their nature of second class, are
denoted again by the letter $\psi$.

Here we notice that sometimes it may be useful to change the matrix
${\bf K}^{12}|_{\Sigma}$ into the unit form. For this we make transformation
\be \label{eq:psi^1-1psi^1}
\;^1\psi_{a_2}^1\;=\;C_{a_2b_2}\psi_{b_2}^1\;,
\ee
then
\be \label{eq:psi^2-1psi^2}
\;^1\dot{\psi}_{a_2}^1\;\stackrel{\Sigma_1}{=}\;C_{a_2b_2}\psi_{b_2}^2\;
\equiv \;^1\psi_{a_2}^2\;.
\ee
Coefficients $C_{a_2b_2}$ will be evaluated from the requirement
$$\;\{^1\psi_{a_2}^1\;,\;^1\psi_{b_2}^2\}\;\stackrel{\Sigma}{=}\delta_{a_2b_2}.$$
We get the transformation matrix:
\be \label{eq:C}
{\bf C}=({\bf F}^{12})^{-1/2}.
\ee

So, it is clear that constraints $\psi_{a_2}^2$ do not generate the tertiary
ones. With taking account of the definition of constraints and the properties
of the Poisson brackets, the property \rf{eq:phi1-phi2-sym} gives rise to
\be \label{eq:PB-psi2-psi2}
\{\psi_{a_2}^2,\psi_{b_2}^2\}~\stackrel{\Sigma}{=}~0.
\ee
Furthermore, with the help of transformation
\be \label{eq:phi-second2-to-1phi-second2}
^1\phi_{\alpha_2}^{m_{\alpha_2}}~=\phi_{\alpha_2}^{m_{\alpha_2}}~+~C_
{\alpha_2~~a_2}^{m_{\alpha_2} m_{a_2}}~\psi_{a_2}^{m_{a_2}}
\ee
one can ensure a realization of the following equality:
\be \label{eq:PB-psi-phi-second2}
\{\psi_{a_2}^{m_{a_2}},\phi_{\alpha_2}^{m_{\alpha_2}}\}~\stackrel{\Sigma}{=}0,
\qquad m_{a_2}=1,2,\quad \alpha_2=1,\cdots,A-A_1-A_2.
\ee
Besides, all the previously established properties
\rf{eq:PB-psi1-psi1} - \rf{eq:PB-psi1-phi-second1} are kept.

Thus at this stage, $A_2$ one-linked chains of second-class constraints are
obtained. In addition, the constraints of different chains are in involution
on $\Sigma$ with each other and with all other constraints.

Next one must consider the constraints $\phi_{\alpha_2}^{m_{\alpha_2}},~~
\alpha_2= 1,\cdots,A-A_1-A_2$, and the matrix
$\left\|\{\phi_{\alpha_2}^1,\phi_{\beta_2}^3\}\right\|$. With the help of the
Jacobi identity we obtain
\be \label{eq:phi1-phi3-antisym}
\{\phi_{\alpha_2}^1,\phi_{\beta_2}^3\}~\stackrel{\Sigma}{=}
~-\{\phi_{\beta_2}^1,\phi_{\alpha_2}^3\}.
\ee
Let
\be \label{eq:rank-A_3}
\mbox{rank} \left\|\{\phi_{\alpha_2}^1,\phi_{\beta_2}^3\}\right\|~
\stackrel{\Sigma}{=}A_3=2R_3<A-A_1-A_2.
\ee
We shall reckon the principal minor of rank $A_3$, disposed in the left upper
corner of this matrix, to be not equal to zero. We consider that
$${\bf K}^{13}=\left\|\{\phi_{a_3}^1,\phi_{b_3}^3\}\right\|~,~~~~a_3,b_3=1,
\cdots,A_3.$$
We have $F_{11}^{13}=0$. Renumbering these constraints we shall attain that
$F_{12}^{13}\neq 0$. We make the transformation
\bea \label{eq:13:phi^1-1phi^1}
\left.\begin{array}{l}
\;^1\phi_1^1\;=\;\phi_1^1\;,\qquad
\;^1\phi_2^1\;=\;\phi_2^1\;,\\
%\quad
\;^1\phi_a^1\;=\;\phi_a^1\;+\;^1u_{a 1}\;\phi_1^1~+\;^1u_{a 2}\;\phi_2^1~,
\quad~~a=3,\cdots,A_3
\end{array}  \right.
\eea
and, hence,
\bea \label{eq:13:phi^3-1phi^3}
\left.\begin{array}{l}
\;^1\phi_1^3\;=\;\phi_1^3\;,\qquad
\;^1\phi_2^3\;=\;\phi_2^3\;, \\
%\quad
\;^1\phi_a^3\;=\;\phi_a^3\;+\;^1u_{a
1}\;\phi_1^3~+\;^1u_{a 2}\;\phi_2^3~, \quad~~a=3,\cdots,A_3.
\end{array}  \right.
\eea
Coefficients $\;^1u_{a 1}\;$ and $\;^1u_{a 2}\;$ are taken as
\be \label{eq:13:1u_a1}
\;^1u_{a 1}\;=\frac{D_{2a}^{13}}{D_{12}^{13}}, \quad \;^1u_{a 2}\;=
-\frac{D_{1a}^{13}}{D_{12}^{13}}
\ee
to satisfy the requirement
$$\;\{^1\phi_1^1\;,\;^1\phi_a^3\}\;\stackrel{\Sigma}{=}0, \quad
\;\{^1\phi_2^1\;,\;^1\phi_a^3\}\;\stackrel{\Sigma}{=}0,
\quad~~a=3,\cdots,A_3.$$
>From here
$$\;\{^1\phi_a^1\;,\;^1\phi_1^3\}\;\stackrel{\Sigma}{=}0, \quad
\;\{^1\phi_a^1\;,\;^1\phi_2^3\}\;\stackrel{\Sigma}{=}0,
\quad~~a=3,\cdots,A_3.$$
Thus we have
$$^1{\bf K}^{13}~\stackrel{\Sigma}{=}\left( \begin{array}{c|c}
{\begin{array}{cc}
0 & F_{12}^{13} \\
-F_{12}^{13} & 0 \end{array}} & {\bf O}\\
\hline
{\bf O} & \|^1{\cal F}_{ab}\|(a,b=3,4,\cdots,A_3)
\end{array}  \right).$$
By continuing this process the matrix ${\bf K}^{13}|_{\Sigma}$ will be
represented in the quasidiagonal form with only nonvanishing elements along
the principal diagonal $F_{a_3b_3}^{1~3}\neq 0$~ (where if $a_3=2k+1,~b_3=
2k+2$ and conversely; $k=0,1,\cdots,A_3-2$).

Again we have the relations:
\bea \label{eq:1-3-chain-properties}
&& \{\psi_{a_3}^2,\psi_{b_3}^2\}~\stackrel{\Sigma}{=}-\{\psi_{a_3}^1,
\psi_{b_3}^3\}, \\
&& \{\psi_{a_3}^2,\psi_{b_3}^3\}~\stackrel{\Sigma}{=}~0,\quad
\{\psi_{a_3}^3,\psi_{b_3}^3\}~\stackrel{\Sigma}{=}~0.
\eea

Thus, two-linked doubled chains of second-class constraints are obtained.
Constraints of such different formations are in involution on $\Sigma$ with
each other and with all other constraints, since all the previously
established properties are kept. Besides one can receive
\be \label{eq:PB-psi-phi-second3}
\{\psi_{a_3}^{m_{a_3}},\phi_{\alpha_3}^{m_{\alpha_3}}\}~\stackrel{\Sigma}{=}0,
\qquad m_{a_3}=1,2,3,\quad \alpha_3=1,\cdots,A-\sum_{1}^{3}A_j
\ee
making the equivalence transformation
\be \label{eq:phi-second3-to-1phi-second3}
^1\phi_{\alpha_3}^{m_{\alpha_3}}~=\phi_{\alpha_3}^{m_{\alpha_3}}~+~C_
{\alpha_3~~a_3}^{m_{\alpha_3} m_{a_3}}~\psi_{a_3}^{m_{a_3}}.
\ee

Turning to the remaining constraints $\phi_{\alpha_3}^{m_{\alpha_3}}$ one must
iterate the above procedure $n$ times. Besides at every $i$-th stage we
consider the constraint set $\phi_{\alpha_{i-1}}^{m_{\alpha_{i-1}}}~(\alpha_
{i-1}=1,\cdots,A-\sum_{1}^{i}A_j)$ with which the constraints chains,
already exhibiting their nature of second class at $i-1$ previous stages of
our procedure, are in involution on $\Sigma$ and suppose that
\be \label{eq:rank-A_i}
\mbox{rank}\left\|\{\phi_{\alpha_{i-1}}^1,\phi_{\beta_{i-1}}^i\}\right\|~
\stackrel{\Sigma}{=}A_i<A-\sum_{1}^{i}A_j.
\ee
Further we have the relation \cite{Chaichian-Martinez-Lusanna}:
\be \label{eq:phi1-phii-isym}
\{\phi_{\alpha_{i-1}}^1,\phi_{\beta_{i-1}}^i\}~\stackrel{\Sigma}{=}
(-1)^i\{\phi_{\beta_{i-1}}^1,\phi_{\alpha_{i-1}}^i\}.
\ee
Renumbering the constraints we obtain that the principal minor in the left
upper corner of the matrix $\left\|\{\phi_{\alpha_{i-1}}^1,\phi_{\beta_
{i-1}}^i\}\right\|$ have the rank $A_i$. Considering it
$${\bf K}^{1i}=\left\|\{\phi_{a_i}^1,\phi_{b_i}^i\}\right\|~,~~~a_i,b_i=1,
\cdots,A_i$$
we see that the matrix ${\bf K}^{1i}$ is (anti)symmetric for (odd) even $i$
on $\Sigma$ (furthermore, its rank is even for odd $i$). After the
(quasi)diagonalization of ${\bf K}^{1i}|_{\Sigma}$ its only nonvanishing
elements are (for odd $i$, $F_{a_ib_i}^{1~i}\neq 0$, where if ~$a_i=2k+1,~b_i=
2k+2$~ and conversely; $k=0,1,\cdots,A_i-2$), for even $i$,~$F_{a_ia_i}^{1~i}
\neq 0,~~a_i=1,\cdots,A_i.$

Furthermore, with the help of the Jacobi identity we have
\cite{Chaichian-Martinez-Lusanna}
\bea \label{eq:1-i-chain-properties}
&& \{\psi_{a_i}^{i-l},\psi_{b_i}^{l+1}\}~\stackrel{\Sigma}{=}(-1)^l
\{\psi_{a_i}^1,\psi_{b_i}^i\},\quad l=0,1,\cdots,i-1, \\
&& \{\psi_{a_i}^j,\psi_{b_i}^k\}~\stackrel{\Sigma}{=}~0,\quad
j+k\neq i+1.
\eea
And also, making the transformation
\be \label{eq:phi-secondi-to-1phi-secondi}
^1\phi_{\alpha_i}^{m_{\alpha_i}}~=\phi_{\alpha_i}^{m_{\alpha_i}}~+~C_
{\alpha_i~~a_i}^{m_{\alpha_i} m_{a_i}}~\psi_{a_i}^{m_{a_i}},\quad
m_{a_i}=1,\cdots,i,~~~\alpha_i=1,\cdots,A-\sum_{1}^{i}A_j,
\ee
one may obtain
\be \label{eq:PB-psi-phi-second}
\{\psi_{a_i}^{m_{a_i}},\phi_{\alpha_i}^{m_{\alpha_i}}\}~\stackrel{\Sigma}
{=}0.
\ee
Thus, at the $i$-th stage we determine $A_i$~ $i-1$-linked chains of
second-class constraints (doubled for odd $i$) which are in involution on
$\Sigma$ with remaining constraints, since all the previously established
properties are kept also.

If after carrying out certain $n$-th stage it is found that
\be \label{eq:rank-0}
\mbox{rank}\left\|\{\phi_{\alpha_n}^{m_{\alpha_n}},\phi_{\beta_n}^{m_
{\beta_n}}\}\right\|~ \stackrel{\Sigma}{=}~0,\qquad
\alpha_n,\beta_n=1,\cdots,A-\sum_{1}^{n}A_j,
\ee
then these remaining constraints $\phi_{\alpha_n}^{m_{\alpha_n}}$ are all of
first class.

So, the final set of constraints $(\Phi,\Psi)$ is obtained from the initial
one $\phi_\alpha^{m_\alpha}$ by the equivalence transformation
\bea \label{eq:Phi-Psi}
\left( \begin{array}{c}
{\bf \Phi} \\
{} \\
{\bf \Psi}
\end{array} \right) = \prod_{a=1}^{X}{\bf S}^a
\left( \begin{array}{c}
{} \\
{\bf \phi} \\
{}
\end{array} \right),\qquad \mbox{det}\prod_{a=1}^{X}{\bf S}^a~\stackrel
{\Sigma}{\neq} 0
\eea
where $X$ is equal to the number of all accomplished stages , ${\bf S}^a$ is
the matrix of the equivalence transformation of each stage.

The total Hamiltonian assumes the final form
\be \label{eq:H_T}
H_T = H + u_\alpha\Phi_\alpha^1,
\ee
where
$$H = H_c+\sum_{i=1}^{n}({\bf K}^{1~i})_{b_i~a_i}^{-1}\{\Psi_{a_i}^i,H_c\}
\Psi_{b_i}^1$$
is a first-class function \cite{Dirac}, $H_c$ is the canonical Hamiltonian,
$u_\alpha$ are the Lagrange multipliers.

Thus, in the Dirac approach, we succeeded in obtaining the canonical set of
constraints with properties analogous to the ones in ref.
\cite{Chaichian-Martinez-Lusanna} without terms quadratic in constraints in
the final form of the total Hamiltonian.

\section{Conclusion}
In the framework of the original generalized Hamiltonian formalism
\cite{Dirac} (without modifications) we have developed a separation scheme of
constraints into the first- and second-class ones on the basis of passing to
an equivalent canonical set of constraints and have determined the general
structure of second-class constraints which is in accordance with the one in
the approaches \cite{Chaich-Mart,Chaichian-Martinez-Lusanna}. The latter has
permitted us to use the classification of constraints and terminology of
paper \cite{Chaichian-Martinez-Lusanna}. That the maximal partition of the set
of constraints is achieved and the canonical set of constraints is obtained,
is seen from that each second-class constraint of the final set has the
vanishing (on the constraint surface) Poisson brackets with all the
constraints of the system except one, and the first-class constraints have the
vanishing Poisson brackets with all the constraints. These precisely
properties will be needed us in following paper II at deriving local-symmetry
transformations.

The important feature of our procedure is that each subsequent stage preserves
the properties of transformed constraints, which were obtained at preceding
stage. This allowed to separate at each stage the second-class constraints.
Note that in the generalized Hamiltonian approach there exists a clear
distinction between primary constraints, which have pure kinematic charakter
as arising only from the definitions of momenta, and the constraints of
subsequent stages of the Dirac scheme for breeding the constraints, which uses
the equations of motion. It was important also (for following derivation of
local-symmetry transformations) to preserve this distinction in the final set
of constraints. Therefore our procedure is constructed so that the secondary,
tertiary, etc. constraints of canonical set do not mix themselves into primary
constraints.

Note that the procedure, proposed in the paper, does not broach questions of
separating functionally-independent constraints satisfying the regularity
conditions. Discussion of these questions can be found in
ref.\cite{Henneaux-Teit}.\\

One of the authors (S.A.G.) thanks the Russian Foundation for Fundamental
Research (Grant N$^{\underline {\circ}}$ 96-01-01223) for support.

%\newpage


\begin{thebibliography}{99}

\bibitem{Dirac}  P.A.M. Dirac, Canad. J. Math. {\bf 2\/}, 129 (1950);
{\it Lectures on Quantum Mechanics\/}, Belfer Graduate School of Science,
Monographs Series, (Yeshiva University, New York, 1964).
\bibitem{Sundermeyer} K. Sundermeyer, {\it Constrained Dynamics\/}, Lecture
Notes in Physics, Vol.169, (Springer - Verlag, Berlin - Heidelberg - New York,
1982).
\bibitem{Gitman} D.M. Gitman and I.V. Tyutin, {\it Canonical quantization of
constrained fields\/}, (Nauka, Moscow, 1986) (in Russian).
\bibitem{Henneaux-Teit} M. Henneaux and C. Teitelboim, {\it Quantization of
Gauge Systems\/}, (Princeton University Press, Princeton, New Jersey, 1992).
\bibitem{Batlle-GPR} C. Batlle, J. Gomis, J.M. Pons and N. Roman-Roy,
J.Math.Phys. {\bf 27\/}, 2953 (1986).
\bibitem{Chaich-Mart} M. Chaichian and D.L. Martinez, Phys.Rev. D {\bf 46\/},
1799 (1992).
\bibitem{Chaichian-Martinez-Lusanna} M. Chaichian, D.L. Martinez and L.
Lusanna, Ann.Phys. (N.Y.) {\bf 232\/}, 40 (1994).
\bibitem{Pavlov} V.P. Pavlov, Theor.Math.Phys. {\bf 92\/}, 451 (1992).
\bibitem{Henneaux-Slavnov} M. Henneaux and A.A. Slavnov,  Phys.Lett. B
{\bf 338\/}, 47 (1994).
\bibitem{Fadd-Jackiw} L.D. Faddeev and R. Jackiw, Phys.Rev.Lett. {\bf 60\/},
1692 (1988).
\bibitem{Sugano-Kimura} R. Sugano and T. Kimura, Phys.Rev. D {\bf 41\/}, 1247
(1990).
\bibitem{Cabo} A. Cabo and P. Louis-Martinez, Phys.Rev. D {\bf 42\/}, 2726
(1990).
\bibitem{Lusanna} L. Lusanna, Riv.Nuovo Cimento {\bf 14\/}, 1 (1991).
\bibitem{GSST} S.A. Gogilidze, V.V. Sanadze, Yu.S. Surovtsev and F.G.
Tkebuchava, {\it The Theories with Higher Derivatives and Gauge-Transformation
Construction\/}, Preprint of Joint Institute for Nuclear Research E2-87-390,
Dubna, 1987; Int.J.Mod.Phys. A {\bf 47\/}, 4165 (1989).

\end{thebibliography}
\end{document}